\documentstyle[prl,aps,amssymb,twocolumn,epsfig]{revtex}

\begin{document}
\draft
\title
{Hall conductance of Bloch electrons in a magnetic field}
\author{D.~Springsguth$^{1,\,2},$ R.~Ketzmerick$^1,$ and  T.~Geisel$^1$}

\address{
$^1$MPI f\"ur Str\"omungsforschung und Institut f\"ur Nichtlineare Dynamik der Universit\"at G\"ottingen, D-37073 G\"ottingen, Germany \\
$^2$Institut f\"ur Theoretische Physik und
SFB Nichtlineare Dynamik, \\
Universit\"at Frankfurt, D-60054 Frankfurt/Main, Germany\medskip\\
\parbox{14cm}{\rm
We study the energy spectrum and the quantized Hall conductance of electrons in a two-dimensional periodic
potential with perpendicular magnetic field {\it without} neglecting  the coupling of the Landau bands. 
Remarkably, even for weak Landau band coupling significant changes in the Hall
conductance compared to the one-band approximation of  Hofstadter's
butterfly are found. 
The  principal deviations are the  rearrangement of
subbands and  unexpected subband contributions to the Hall conductance.
\smallskip\\
}}

\maketitle
\narrowtext

\section{ Introduction}

Since many decades the problem of electrons under the
influence of a two-dimensional periodic potential (Bloch electrons) and a perpendicular magnetic field is of great interest \cite{pei}. 
Each of the limiting cases, just a periodic potential and just a magnetic
field, was solved in the early days of quantum mechanics \cite{lndau,bloch}. Their solutions are translation invariant Bloch 
waves with energy {\it bands} and rotation invariant oscillator functions with
discrete Landau {\it levels}, respectively. Away from the limiting cases the system must combine these adverse properties.   
For very  weak  and for very strong
magnetic fields, compared with the potential strength, this combination
gives rise to a fractal energy spectrum -- the famous Hofstadter butterfly \cite{hofst} (see Fig.~2). 
  It is based on a one-band approximation that leads to the tight-binding
  Harper equation \cite{harp,rauh}. In the intermediate regime, where the magnetic
  field is  of comparable strength to the potential, one has to take into account 
  the coupling between the Landau Levels or between the Bloch bands. 
 In doing so, one obtains a vectorial tight-binding equation\cite{pets},
 which has the correct chaotic classical limit\cite{pets,geis}, whereas the one-band approximation of the Harper
 equation has  an integrable classical limit. 
The coupling causes  considerable changes in the Hofstadter butterfly and
is of importance for experimental observations \cite{pets,kuehn,silber}.

Currently, one tries to find signatures of Hofstadter's butterfly in lateral
superlattices with periods of about $100\,nm$ on GaAs-AlGaAs heterojunctions.
Straightforward spectroscopic measurements are not yet feasible \cite{gudm} and, instead, the
efforts are concentrated on 
 magnetotransport measurements \cite{weiss,enss}. 
 In fact, substructure in the Shubnikov-de Haas oscillations (oscillations
  in the longitudinal resistance due to the Landau levels) was 
  found, which demonstrates the splitting of Landau levels due to 
  the periodic potential \cite{enss}.

\begin{figure}[h]
\hspace*{-1cm}
\epsfig{figure=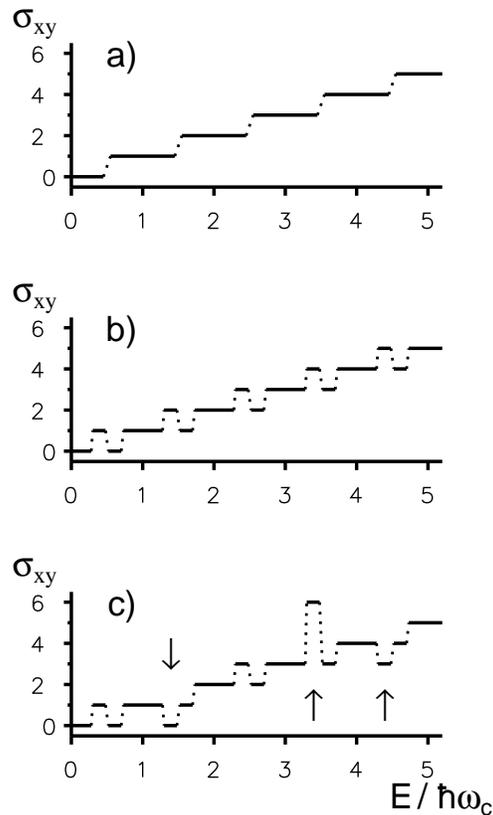,width=7.3cm}
\caption{\footnotesize
The Hall conductance (solid lines) in the energy gaps in units of $( e^2/h)$ is plotted schematically versus energy in units of cyclotron energy $\hbar\omega_{c}$ 
in the cases a) without a periodic potential (quantum Hall effect), b) with a periodic potent ial neglecting coupling of Landau bands, and c) including coupling of Landau bands (see also Fig.~3b).  
The magnetic flux per unit cell is $3/2$. The dotted lines serve as  a guide to the eye. One can see that the coupling can dramatically change the Hall conductances (arrows). }
\end{figure}
  Beyond these qualitative findings in the longitudinal resistance the
  study of the Hall conductance would provide a quantitative demonstration of
  the Landau level substructure:
Von Klitzing et al.\   discovered in 1980 that the Hall conductance $\sigma$ is
 quantized between Landau levels in integer multiples of $ e^2/h$ \cite{klit}
 (Fig.~1a). 
 Under the influence of a weak periodic potential each Landau level 
 broadens into a Landau band  with so-called minigaps and 
one might have thought that the Hall conductances in these minigaps  were 
rational multiples of $e^2/h$.   Thouless et al.\ \cite{thou}, however,  showed with an argument by
Laughlin \cite{laugh},  that the Hall conductance is quantized even in these minigaps  
  in {\it integer} multiples of $e^2/h$. 
 For Hofstadter's butterfly they found that these integer values vary irregularly from gap to gap according to a diophantic equation 
 (see Figs.~1b and 2) \cite{hatsu}. 
 Whereas
 the longitudinal resistance is zero in every gap, the Hall conductance
 differs from gap to gap and thus contains quantitative  information about
 the Landau band substructure. 

In order to make the minigaps observable in the presence of finite
disorder broadening \cite{huck}, one has to sufficiently increase the potential
strength. This increases the width of the 
  Landau bands  and the minigaps, but at the same time increases the
  coupling between the Landau bands. 
This coupling, however, changes the structure of the
energy spectrum considerably \cite{pets}, and the results for the Hall
conductance of Hofstadter's butterfly  do no longer apply (Fig.~1c). 
The integer
quantization of the Hall conductance in any gap, on the other hand, is  
 ensured by Laughlin's argument even in this general
case. 
Therefore the
question arises: How will the Landau band coupling influence the integer values of the Hall conductance? 
We will answer this question by studying the energy spectrum and the Hall
conductance for different
strengths of the Landau band coupling. 
The Hall conductance in a gap  can change only if the gap closes and reopens as a
function of the coupling strength. 
Surprisingly, this happens even for weak Landau band coupling and we find the following principal deviations from the Hall conductance in  Hofstadter's butterfly: i) opening of previously closed gaps, ii) rearrangement of
subbands, including their contributions to the Hall conductance, and iii) unexpected subband contributions to the Hall conductance.
We finally make some remarks on the observability in lateral superlattices
on semiconductor  heterojunctions. 

In the last years, electrons in two dimensions have been studied also under the influence of a
magnetic modulation \cite{niu}, instead of an electric modulation. It was shown that there exists
a connection between these two cases, e.g.\  
 the Hall conductance  in systems with magnetic modulation is also quantized in an energy gap. 
 Consequently, the phenomena discussed in this paper may also apply to these systems.

In Sec.~II the model is introduced. The well known Hall conductances when neglecting the Landau
band coupling are 
presented in Sec.~III. In Sec.~IV we study the influence of Landau band coupling on the energy
spectrum and on the quantized values of the Hall conductance. 
The experimental observability of these findings is discussed in Sec.~V and Sec.~VI gives
concluding remarks.

\section{ Model}

The one-particle Hamiltonian for an electron with charge $-e$ and effective mass $ m^{\ast}$ in a magnetic field and in a two-dimensional potential has the form  
\begin{equation} \label{hamil}
H  =\frac{1}{2m^{\ast}}({\bf p}+e{\bf A})^{2}+V(x,y)~, 
\end{equation}
where  we neglect  spin  and electron-electron interaction. 
In a homogenous magnetic field {\it \bf B} in $z$-direction  the vector potential  in the Landau gauge 
is  given by {\it {\bf  A}}$=B(0,\,x ,\,0)\,$ and 
the periodic potential can be written in its Fourier decomposition  
\begin{equation}\label{fourier}
V(x,y)=V_{0}\sum_{r,s} v_{r,s}~e^{2\pi i(rx/a+sy/b)}\,,
\end{equation}
with $a$ and $b$ the  periods in  $x$- and $y$-direction, respectively, and $V_{0}$ the difference between maximum and minimum of the
potential. 

We choose as a basis the product ansatz of the limiting solutions in
coordinate representation,
namely  plane waves and oscillator functions
\begin{equation}
\langle x,y\;|\;\nu,\theta \rangle=N~e^{i\frac{y}{b}\theta}~~\Psi_{\nu}(\mbox{$\frac{x}{l}+\frac{\theta l}{b}$})~,
\end{equation}
where $l=\sqrt{\hbar/(e\,B)}$ is the magnetic length,  $\Psi_{\nu}(z)=exp(-z^2/2)~H_{\nu}(z)$ with $H_{\nu}$ the $\nu-$th Hermite polynomial, normalization $N$ \cite{pfann}, $\nu =0,\,1,\,2,\ldots\,$, and $\theta 
 \in \mathbb{R}$. 
The matrix elements of the first part of the Hamiltonian read  
 \begin{eqnarray}
\langle \mu,\varphi\;|\;\frac{1}{2m^{\ast}}({\bf p}+e{\bf A})^{2}\;|\;\nu,\theta \rangle=\hbar\omega_{c}~(\nu+1/2)~\delta_{\mu,\nu}~\delta_{\varphi,\theta}~,
\end{eqnarray}
which are the Landau levels with  cyclotron frequency
$\omega_{c}=eB/m^{\ast}$, and those of the potential
\begin{eqnarray}
\langle \mu,\varphi\;|v_{r,s}\,e^{2\pi i(rx/a+sy/b)}|\nu,\theta \,\rangle=&&\nonumber\\ 
P_{\mu \nu}(r,s)~&e^{-i\frac{\Phi_{0}}{\Phi}r\,\varphi}~&\delta_{\varphi,\theta +2\pi\,s}
\end{eqnarray}
with 
\begin{eqnarray}\label{pnumu}
 P_{\mu \nu}(r,s)&=&v_{r,s}~e^{irs\pi \frac{\Phi_{0}}{\Phi}}~e^{-\frac{u}{2}}~\mbox{$\sqrt{\frac{\nu !}{\mu !}}$} ~(\pi\,{\textstyle \frac{\Phi_{0}}{\Phi}})^{\frac{\mu- \nu}{2}}\nonumber\\
 &&\times(s\alpha^{-1}+ir\alpha)^{\mu -\nu}~L^{\mu -\nu}_{\nu}(u), \ \ \ \quad \mu \geq \nu
\end{eqnarray}
and 
\begin{eqnarray}\label{pmunu}
 P_{\mu \nu}(r,s)&=&v_{r,s}~e^{irs\pi \frac{\Phi_{0}}{\Phi}}~e^{-\frac{u}{2}}~\mbox{$\sqrt{\frac{\mu !}{\nu !}}$} ~(\pi\,{\textstyle \frac{\Phi_{0}}{\Phi}})^{\frac{\nu- \mu}{2}}\nonumber\\
 &&\times(-s\alpha^{-1}+ir\alpha)^{\nu -\mu}~L^{\nu -\mu}_{\mu}(u), \ \ \ \quad \nu \geq \mu
\end{eqnarray}
 $L^{\nu}_{\mu}(u)$ the Laguerre polynomials, $\alpha =\sqrt{b/a}$, and  $u=\left[ \pi\, (r^{2}\alpha^2\,+\,s^2 \alpha^{-2})\,\Phi_{0}/\Phi  \right]$\,.
Here $\Phi_{0}/\Phi$ is the ratio of the magnetic flux quantum
$\Phi_{0}=h/e$ divided by the flux $\Phi$  through a unit cell of the 
 periodic potential 
\begin{equation}\label{f}
\frac{\Phi_{0}}{\Phi}=\frac{h/e}{a\,b\,B}~.
\end{equation}

We divide the parameter $\theta$ into  $\theta=2\,\pi\,n+\vartheta$ with integer $n \in \mathbb{Z}$ and phase $\vartheta \in [0,2\pi)$, as the Hamiltonian is diagonal in
$\vartheta$.  
The eigenstates $|j, \vartheta\rangle$ of the Hamilton operator are decomposed into the basis states by 
\begin{equation}
 \label{koeff} |j,\vartheta\rangle=\sum_{\nu,n}a^{\nu}_{n}(j,\vartheta)~|\nu,n,\vartheta \rangle .
\end{equation}
Inserting this into the Schr\"odinger equation 
 one obtains for every $\vartheta$   the following  eigenvalue equation   
\begin{equation}
\label{schr}
{\bf A}_{n}~a_{n}~+~ \sum_{s \not = 0} {\bf T}_{n,s}~a_{n+s} =
\frac{E}{\hbar\omega_{c}}~a_{n}
\end{equation}
with the vector $a_{n}$
\begin{equation}
a_{n} =(a^{0}_{n},a^{1}_{n},\ldots,a^{\nu}_{n},\ldots),
\end{equation}
and the matrices ${\bf A}_{n}$ and ${\bf T}_{n,s}$~,
\begin{eqnarray}
{\rm A}^{\mu \nu}_{n}&=&(\nu+1/2)\delta_{\nu,\mu}\nonumber\\
&&+K\cdot\frac{\Phi_{0}}{\Phi}\sum\limits_r P_{\mu \nu}(r,0)~e^{-i\, r(2\pi n+\vartheta) \frac{\Phi_{0}}{\Phi}}~, \label{anu}\\
\nonumber\\
{\rm T}^{\mu \nu}_{n,s}&=&K\cdot\frac{\Phi_{0}}{\Phi}\sum\limits_r P_{\mu \nu}(r,-s)~e^{-i\, r(2\pi n+\vartheta) \frac{\Phi_{0}}{\Phi}}  \label{tnu}~.
 \end{eqnarray}
Here the important parameter 
\begin{equation}\label{KK}
K=2 \pi\, m^{\ast}\,a\,b\, V_{0}/h^2
\end{equation}
is a measure for the
 strength of the coupling of the Landau bands. 

Equation~({\ref{schr}}) is an infinite dimensional matrix equation, which cannot
be diagonalized 
numerically. 
Only if $\Phi_{0}/\Phi$ is a rational number 
\begin{equation}\label{rational}
 \frac{\Phi_{0}}{\Phi}=\frac{h}{e\,B\,a\,b}=\frac{p}{q}\,, \mbox{with~} p,q \in \mathbb{N}~,
 \end{equation}
 which means that $q$ flux quanta penetrate $p$ unit cells, 
 one can make use of the magnetic translation operators \cite{kohm,uso}. 
 For the vector potential in the Landau
 gauge  they are defined by 
 \begin{equation}
 M_{a}=e^{i\,ya/l^2}~e^{a\partial_{x}}~\mbox{and}~M_{b}=e^{b\partial_{y}}~,
 \end{equation}
 and displace a wave function by one unit cell in $x-$ or $y-$direction. 
 The magnetic translation operators commute with the Hamilton operator, but
 in general not with each other. 
 Only in the case that $\Phi_{0}/\Phi$ is a rational number (Eq.~(\ref{rational})), one can  
 enlarge the unit cell of the periodic potential by 
 a factor of $p$ to a new magnetic unit cell 
  and finds 
 \begin{equation}
 [M_{pa},M_{b}]=0~.
 \end{equation}
 Then the eigenfunctions of the Hamilton operator are also  eigenfunctions of
 $M_{pa}$ and $M_{b}$   
with eigenvalues $e^{i\kappa}$ and $e^{i\theta}$, respectively, 
and 
$\kappa \in [0,2\pi)$ and $\theta \in [0,2\pi q)\,$.   
It follows 
 \begin{equation}
 a^{\mu}_{n-q}(j,\vartheta,\kappa)=e^{i\kappa}a^{\mu}_{n}(j,\vartheta,\kappa)~,
 \end{equation}
  so that $n$ can be restricted to $0,1,\ldots,q-1$, producing $q$ subbands. 
 Furthermore, if  we take only $N$ consecutive Landau bands into account, Eq.~({\ref{schr}}) 
 reduces to a finite  $(Nq\times Nq)$ eigenvalue equation for every $\kappa$ and every $\vartheta$.
As one can see from Eqs.~(\ref{anu}) and (\ref{tnu}), the energy spectrum in units of $\hbar \omega_{c}$ for a given potential shape  depends 
only on the number of flux quanta per unit cell $\Phi_{0}/\Phi$ and the
Landau band coupling $K$. 
 
 \section{Hall conductance without Landau band coupling}

If one  neglects the coupling of
the Landau bands, i.e. neglects the terms with $\nu \not = \mu$ in Eqs.~(\ref{anu}) and (\ref{tnu}), and takes into account only the lowest Fourier components
(Eq.~(\ref{fourier})), i.e. the $\cos x +\cos y$-potential 
\begin{equation}
\label{simple}
V(x,y)=\frac{V_{0}}{4}~\Big(\cos(\mbox{$\frac{2\pi x}{a}$})+\cos(\mbox{$\frac{2\pi y}{a}$})\Big )~,
\end{equation}
one obtains  the Harper equation \cite{harp}
\begin{equation}
a_{n+1}~+~a_{n-1}~+~2\cos\left[(2\pi n+\vartheta)\frac{\Phi_{0}}{\Phi}\right]a_{n}=\tilde{E}\, a_{n}~,
\end{equation}
with the scaled energy 
\begin{equation} \label{Etilde}
\tilde{E}=\Big(E-\hbar\omega_{c}\,(\nu +1/2)\,\Big) /\Big(P_{\nu,\nu}(1,0)\,V_{0}\Big)~.
\end{equation}
For every Landau band $\nu$ the resulting energy subbands plotted against the inverse flux $p/q$ show the well-known Hofstadter butterfly \cite{hofst}
 (see Fig.~2). 

The Hall conductance for this case was derived by Thouless et al. \cite{thou} as the solution of a diophantic equation. In
units of $e^2/h$ the Hall conductance $\sigma$  in the $g-$th gap of Hofstadter's butterfly is given by   
\begin{equation}
\label{thoul}
g=w\,p\,+\,\sigma\,q\hspace*{1cm} \mbox{with}~ |w|\leq q/2\,,
\end{equation}
$g,p,q \in \mathbb{N}$, and $w,\sigma \in \mathbb{Z}$ \cite{semiclass}. 
Figure~2 shows the Hofstadter butterfly with the Hall conductance of the lowest Landau band written in the large gaps. 
For the Hall conductance in higher Landau bands one has to add the Landau band index $\nu$ to
these values. 
Figure~2 shows  that the Hall conductance is not always  a monotonous function of energy,  as in the case without
potential (Fig.~1a), a fact that could be the first experimentally obtainable hint for the internal band structure. 
But will this figure remain valid if  the coupling of the Landau bands is taken into account? 

\vspace*{0.5cm}
\begin{figure}[h]
\hspace*{-1cm}
\epsfig{figure=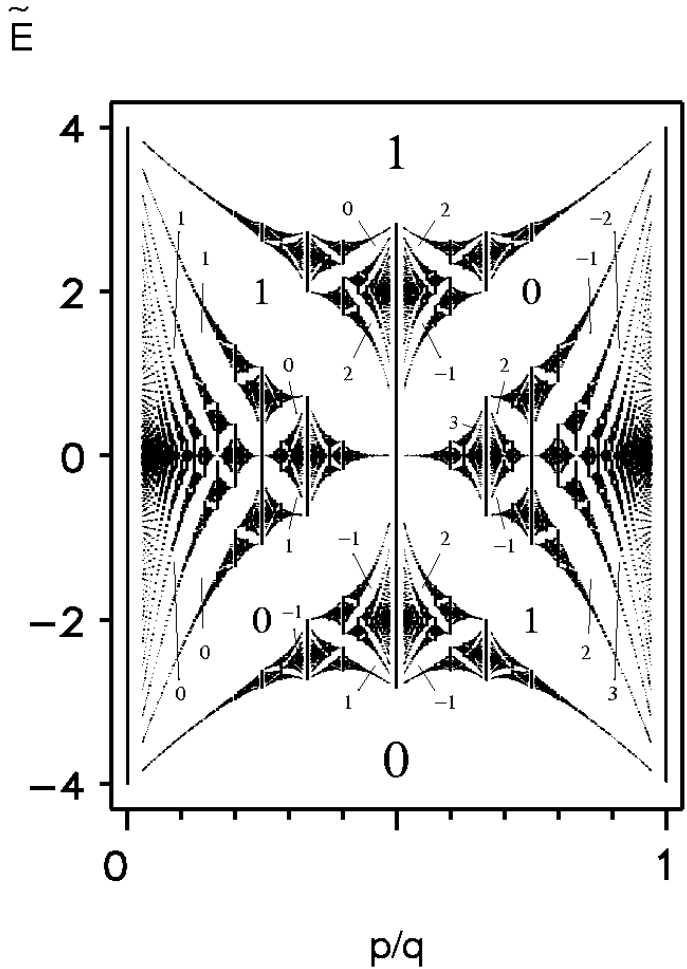,width=9.3cm}
\vspace*{0.8cm}
\caption{\footnotesize
Neglecting the Landau band coupling the scaled energy $\stackrel{\sim}{E}$ (Eq.~(\ref{Etilde})) versus the inverse magnetic flux $p/q$ yields 
for e very Landau band the same spectrum, namely the Hofstadter butterfly. The numbers in the energy gaps are the quantized Hall con ductances in units of $(e^2/h)$ to 
which one has to add the Landau band index $\nu$.}
\end{figure}

\section{Hall conductance with Landau band coupling}
The influence of the coupling between the Landau bands on the spectrum is determined by the coupling strength 
$K$ (Eq.~(\ref{KK})). 
How does an increase  of the coupling strength affect the spectrum? 

Figure 3 shows the energy spectrum \cite{zz} as a function of $p/q$ for three different values of the coupling strength $K$ for the
$\cos x+\cos y$-potential~(\ref{simple}). 
For small $K$ (Fig.~3a) each Landau band resembles the Hofstatder butterfly multiplied
with $P_{\nu,\nu}$ (Eq.~(\ref{Etilde})). The Laguerre polynomials in
this expression become zero for certain $\Phi_{0}/\Phi$, so that the width of the corresponding Landau band vanishes, the  so-called flatband positions.  
Their number increases with the Landau band index. 
With increasing $K$ each Landau band becomes  wider, even at the original ($K \ll 1$) flatband
positions, and more distorted (Fig.~3b). 
For $p/q$ with even $q$ the $q/2$-th minigap, which is 
 closed without coupling, opens due to the coupling. 
If the coupling is strong enough, Landau bands may even overlap. 
In this case the classification into Landau bands becomes meaningless (Fig.~3c). 
Similar effects occur  also for other periodic potentials, e.g. for potentials of
the form
\begin{equation}
\label{potenz}
V(x,y)=V_{0}\Big(\cos(\mbox{$\frac{\pi x}{a}$})\,\cos(\mbox{$\frac{\pi y}{a}$})\Big)^{\beta}
\end{equation}
which are used as  model potentials for antidots \cite{flei}(see Fig.~5b).
Such spectra for Bloch electrons in a magnetic field with Landau
band coupling were previously studied in Refs.~\cite{pets} and \cite{kuehn},
 where slight mistakes in the matrix elements of the potential, however, led to 
 different results (in Ref.~\cite{kuehn} only for antidot potentials
 (Eq.~(\ref{potenz})) with $\beta >2$).

The diophantic equation~(\ref{thoul}) for the Hall conductance is 
valid for the general case with coupling  and arbitrary periodic potential only without the constraint
$|w|\leq q/2$ \cite{dana,avron}, however, it then allows many solutions for the Hall
conductance in a given gap. Instead, one
may take advantage of a formula derived by St\v{r}eda \cite{streda}: 
For any energy gap the Hall conductance (in units of $e^2/h$) is given by  
\begin{equation}\label{stred1}
\sigma=\frac{\partial N(E)}{\partial B}\frac{h}{e}
\end{equation}
with $N(E)$ the number of states per unit area having energy lower than the gap energy. 
As the Hall conductance in the gap is quantized, it is possible to 
replace $\partial N/\partial B$ by $\bigtriangleup N/\bigtriangleup B$ for
adjacent rationals $p_{1}/q_{1}$ and $p_{2}/q_{2}$ which share this gap. 
Using the fact that
 the number of states per unit area in a subband is $eB_{i}/h\,q_{i}\,(i=1,\,2)$ and Eq.~(\ref{rational})  one finds  
 \begin{equation}
 \sigma =\Big(\frac{n_{1}}{p_{1}}-\frac{n_{2}}{p_{2}}\Big)/\Big(\frac{q_{1}}{p_{1}}-\frac{q_{2}}{p_{2}}\Big)=\frac{n_{1}p_{2}-n_{2}p_{1}}{q_{1}p_{2}-q_{2}p_{1}}~, \label{streda}
 \end{equation}
where $n_{i}$ is the number of subbands below the gap for the corresponding magnetic
field $B_{i}$. 
The Hall conductance in a gap can only change if the gap closes and opens again as a function of
the Landau band coupling \cite{tesa,niu}. 
We find three types of deviations from the Hall conductance of Hofstadter's butterfly:
i) opening of previously closed gaps, ii) rearrangement of
subbands, including their contributions to the Hall conductance, and iii) unexpected subband contributions to the Hall conductance.
We will now discuss these deviations in more detail. 
\clearpage
\begin{figure*}[t]
\centering
\epsfig{figure=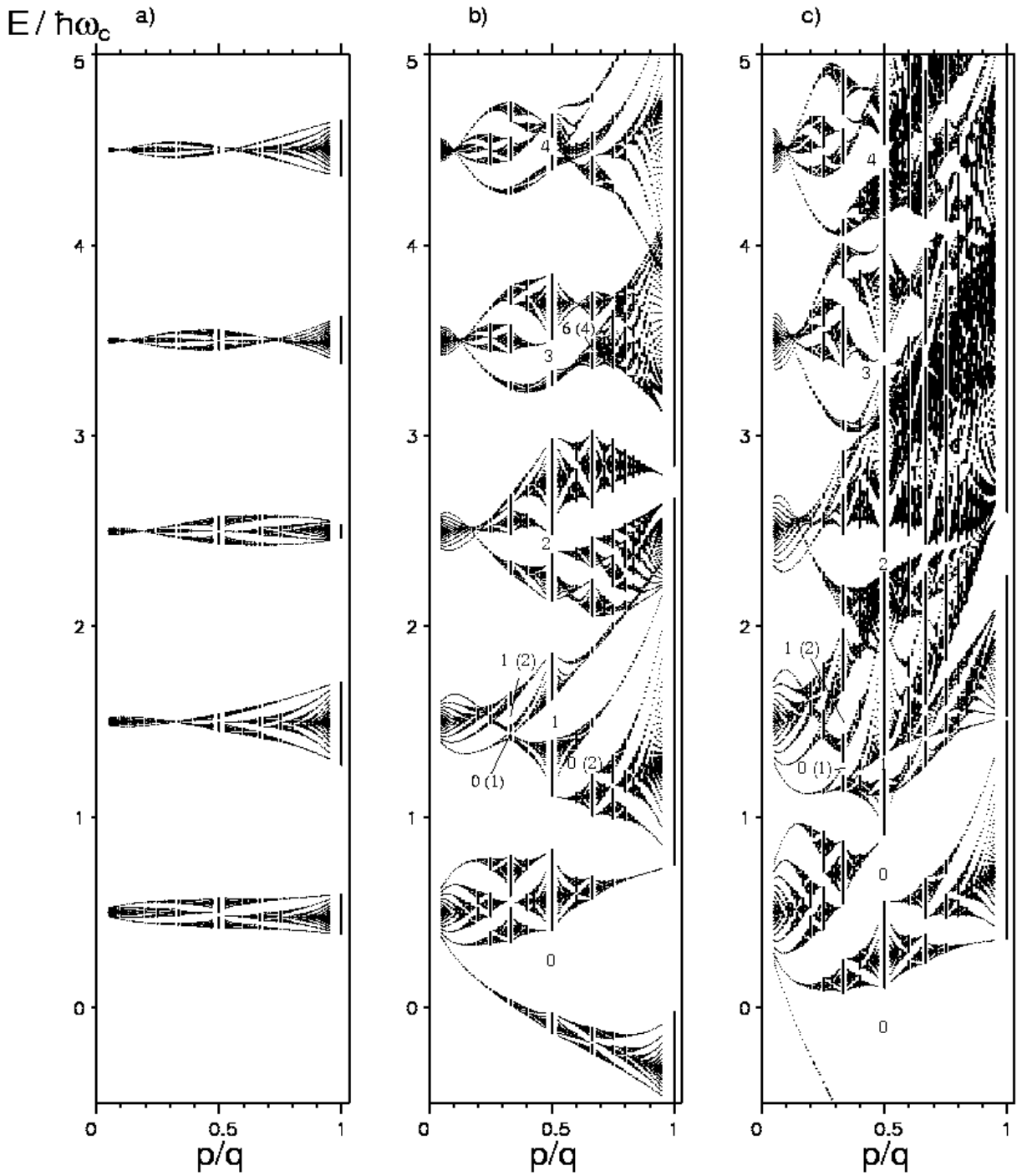,width=16.5cm}
\vspace*{1.9cm}
\begin{minipage}[b]{17cm}
\vspace*{1cm}
\caption{\footnotesize
The five lowest Landau bands for the $\cos x+\cos y$-potential (Eq.~ (\ref{simple})) are plotted for increasing coupling strength $K=1,\,6,\,12$. 
For $p/q=1/2$ the Hall conductances in the minigaps are shown and also for $p/q =,1/3,$ and $2/3$, if 
they  deviate from the corresponding value of Hofstadter's butterfly (given in brackets). 
 The number of Landau bands considered numerically  is 6, 9, and 13, respectively.} 
 \end{minipage}
\end{figure*}
\clearpage

i) The  gaps in the middle of a Landau band for $p/q$ with $q$ even are closed in Hofstadter's butterfly (Fig.~2), but due to the coupling of Landau bands they may open (Fig.~3). Consequently, there
appears a new
plateau in the Hall conductance. As an 
aside, we note that near flatband positions, where a Landau band is small, the minigap for
$p/q=1/2$ is often surprisingly large (see the 11th and 14th Landau band in Fig.~5a). 

ii) The second effect occurs, e.g.\  at $p/q=1/3$ in Fig.~3b (see also Fig.~4). 
Without coupling one would expect the Hall conductance in the second minigap of the second Landau band to be  2, namely 1 from Hofstadter's butterfly (Fig.~2) plus 1 from the one Landau band below.  
 With coupling, instead, we find Hall conductance 1. 
One can  understand this effect by looking at the entire spectrum for a
given $K$ (Fig.~3): 
With increasing $p/q$ the  uppermost (for small $p/q$) branch of the second Landau band is bent downward
 in such a way, that at $p/q=1/3$ the  middle branch lies at the top of the Landau band and so does its contribution (+1) to the  Hall conductance. 
 This rearrangement gives rise to the change of the Hall conductance from 2
 to 1 in the second minigap. It has consequences also for $p/q=2/3$ in the second Landau band. 
There the sequence of the Hall conductances in the minigaps reads, starting below the
Landau band,  
$\sigma=\{1,0,1,2\}$ instead of $\{1,2,1,2\}$. The subband carrying the Hall conductance (+1)
is exchanged with the  one carrying (-1). 

\begin{figure*}[b]
\centering
\epsfig{figure=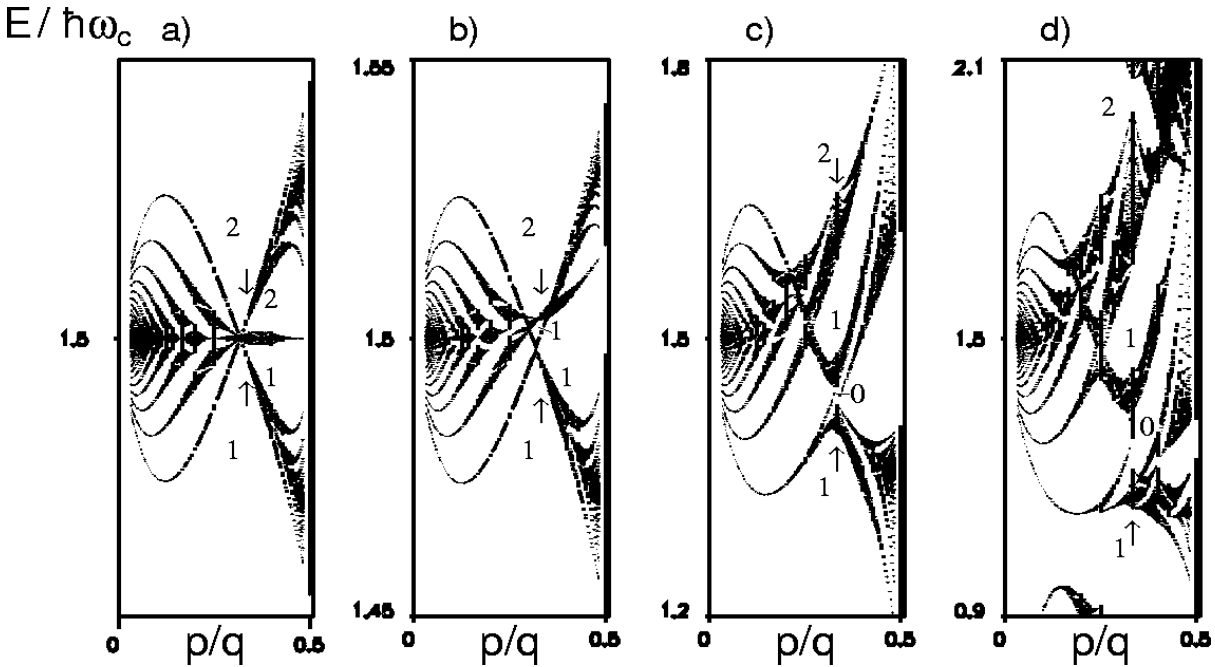,width=16.0cm}
\vspace*{0.3cm}
\begin{minipage}[t]{17cm}
\caption{\footnotesize
The second lowest Landau band of the  $\cos x+\cos y$-potential (Eq.~(\ref{simple})) is shown a) without, b) with weak ($K=1$), c) moderate ($K =6$), and d) 
strong coupling of Landau bands ($K=12$). One can see that even for a very weak periodic potential there can exist deviations from the diophantic 
equation~(\r ef{thoul}) near flatband positions, here at $p/q=1/3$ (arrows) as discussed in the text.}
\end{minipage}
\end{figure*}
\newpage

iii) A surprising effect is found in Fig.~3b (see also  Fig.~1c).  
 Applying the diophantic equation~(\ref{thoul}) to the case $p/q=2/3$ each subband carries a
Hall conductance $(+1)$ or $(-1)$. But in the 4th Landau band the sequence of the
plateaus is found to be $\sigma=\{3,6,3,4\}$, which means that two of the subbands contribute instead with $(+3)$ and $(-3)$, respectively. 
Something similar happens in  Figs.~3b and c for $p/q=1/3$: Without the coupling of the Landau bands one expects for the second Landau band a  monotonous sequence $\sigma=\{1,1,2,2\}$, with  coupling it reads $\{1,0,1,2\}$, now the lowest subband carries a negative conductance.  
These two examples are in full agreement with a formula derived by Dana et al.
\cite{dana}, which gives  all  possible contributions of subbands to the Hall
 conductance for any
periodic potential, namely  
\begin{equation}
1=m\,p~+~\bigtriangleup \sigma\, q~,
\end{equation}
 where $\bigtriangleup \sigma$ is the contribution of a subband and $m \in \mathbb{Z}$. 

 But it still leaves the question:
 How do such unexpected contributions arise? In Fig.~2 one sees, that for
 $p/q=2/3$ 
  without coupling there exists a minigap with a Hall conductance (+3) to the left of $p/q=2/3$, which ends at the second subband of $p/q=2/3$. In
 fact, as a function of the coupling strength the first minigap for $p/q=2/3$ in the 4th Landau band 
 closes and opens in such a way, that now it includes  the minigap with Hall
 conductance (+3), giving rise to the sequence $\sigma=\{3,6,3,4\}$. 
 \newpage
 Similarly, one can explain the above example at $p/q=1/3$, where one finds a 
minigap to the left of the first subband of $p/q=1/3$ with the Hall
conductance (-1) (Fig.~2).

Remarkably, even for very weak Landau band coupling we found examples for subband rearrangement (ii).  
For vanishing coupling and for the $\cos x+\cos y$-potential~(\ref{simple}) the width of the Landau band at flatband positions  is zero. 
For example, in the  second Landau band near $p/q=1/3$ the bottom, middle and top branch of the band
cross in one point (Fig.~4a). 
This degeneracy is lifted, as soon as the Landau band coupling is turned on,
and subbands are rearranged, including their contribution to the Hall
conductance. The Hall conductance in the second Landau band at $p/q=1/3$
thus changes from $\sigma=\{1,1,2,2\}$ to $\{1,1,1,2\}$ (Fig.~4b). 
 With increasing coupling strength the $p/q$-range of this rearrangement expands (Fig.~4c). 
Furthermore, one finds in Figs.~4c and d the Hall conductance  0 (instead of
1) in the first minigap of $p/q=1/3$, discussed above as an example of
deviation iii). It has its origin in the correspoding crossing of branches
in Fig.~4b. 

Another example of a  deviation from the diophantic  equation~(\ref{thoul})
even for weak Landau band coupling can be  found in the 4th Landau band near $p/q =1$ for ratios of the form $(q-1)/q$. 
Each subband adds  one unit of conductance, except for the middle one that carries the  large negative
conductance (2-q), so that the sum of the contributions for a Landau band equals one. 
Specifically, for $p/q=6/7$ the sequence of the values of the Hall conductance in the
gaps between the subbands for the 4th Landau band reads without coupling with
increasing energy 
\begin{equation}
\sigma=\{3,4,5,6,1,2,3,4\}~,\nonumber
\end{equation}
as can be seen in Fig.~2 with (+3) added for the lower three Landau bands. 
Even for weak coupling (Fig.~3a) one finds instead  
\begin{equation}
\sigma=\{3,4,5,0,1,2,3,4\}\nonumber~,
\end{equation}
 where  the step to lower values occurs earlier in the sequence. 
One can interpret this effect by the exchange of  the subband carrying (-5) with one carrying (+1). 
Again, it is due to the sensitivity of flatband positions to the Landau
band coupling. 

Figure 5 compares a sequence of 15 Landau bands for the $\cos x+\cos y$-potential~(\ref{simple}) and the antidot potential~(\ref{potenz}) and shows
 the deviations from the Hall conductances of
Hofstadter's butterfly for $p/q=1/2,1/3,$ and $2/3$. For the antidot potential one finds even without Landau band coupling 
 several deviations 
 from the diophantic equation~(\ref{thoul}) due to the different potential shape. 
 Taking the coupling into account, we find
 the three principal types i)-iii) of deviations from the case without coupling  as discussed
 above.
Again, even weak Landau band coupling gives rise to these deviations near
crossings (as a function of $p/q$) of branches of the spectrum, which are
strongly affected by a weak coupling.

To summarize, we find that crossings of branches of the spectrum, e.g.\
at flatband positions, may lead to
deviations of the Hall conductance from the diophantic equation~(\ref{thoul}) even for weak 
coupling (Fig.~4). With increasing coupling the $p/q$-range of these deviations
expands.

\section{Remarks on Observability}

Bloch electrons in a magnetic field may be experimentally studied in
lateral superlattices on semiconductor heterojunctions.  The main obstacle for observing the
subband structure is that most minigaps are small 
compared to the disorder broadening. 
A crude estimation of the disorder broadening can be given in the self-consistent
Born approximation \cite{ando}:  
A single Landau level (or one of the $q$ subbands) will be broadened 
by disorder 
 to a sharp half-ellipse with a total width of $2\,\Gamma$ (or $2\,\Gamma/\sqrt{q})$, with $\Gamma$  given by 
\begin{equation}
\Gamma^{2}~=\frac{2}{\pi}~ \frac{\hbar}{\mu~ m^{\ast}/e}~\hbar
\omega_{c},
\end{equation}
with the mobility $\mu$. 
For typical values $\mu=50\,m^2/(Vs), a=100\,nm$, and
 $m^{\ast}=0.067\,m_{e}$ this equation gives $\Gamma=0.18\,\sqrt{p/q}\,\hbar\omega_{c}.$ 
All  the minigaps in Fig.~3a and all, except for the largest ones,  in Fig.~3b
  are closed by such a  disorder broadening. 

Thus, for a given disorder broadening one has 
to increase
the  strength of the potential in order to enlarge the internal gaps, 
 using the fact that the width of a Landau band increases proportional to $V_{0}$. 
Increasing the potential strength, however,  
  also  increases the coupling strength, which influences the spectrum.
 If the coupling is too strong, the Landau bands  merge, many gaps are closed
 and it is  difficult to interpret the spectra in terms of Landau bands (Fig.~3c).  
One has to choose the coupling strength $K$ in such a  way, that the Landau bands  are as wide as possible in order to obtain observable gaps, while not  overlapping with adjacent bands   in the desired range of $\Phi_{0}/\Phi$. 
A quantitative estimation beyond $V_{0}=\hbar\omega_{c}$, which is
equivalent to $K=\Phi/\Phi_{0}$, would have to consider that the shape of the Landau
bands differs from band to band, and that it depends strongly on 
strength and Fourier components of the potential, 
as can be seen in Fig.~5 . 
\clearpage
\begin{figure*}[t]
\centering
\epsfig{figure=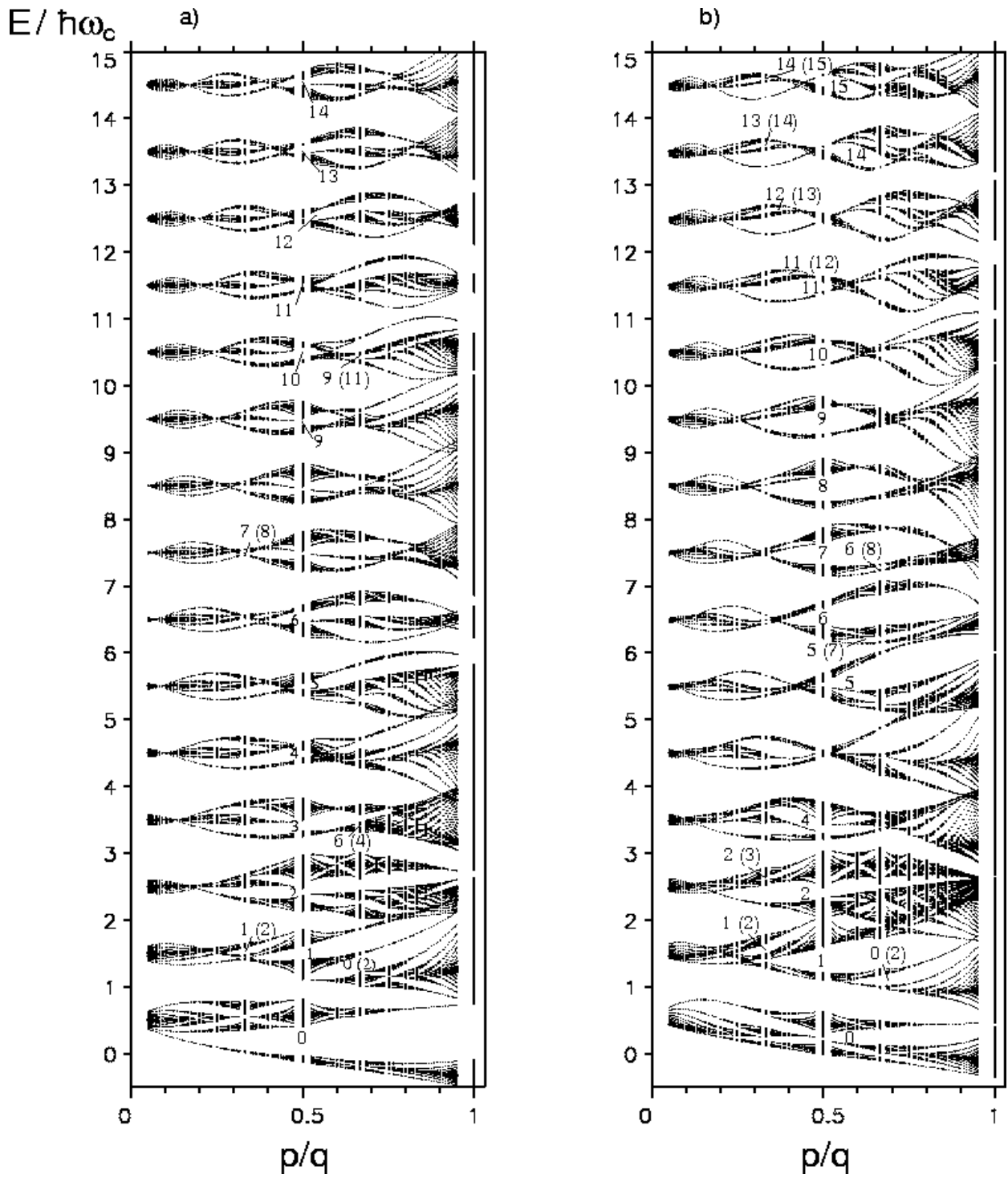,width=16.5cm}
\vspace*{1.3cm}
\begin{minipage}[t]{17cm}
\caption{\footnotesize
The lowest 15 Landau bands are plotted for $K=6$ for the $\cos x+\cos y$-potential (Eq.~(\ref{simple})), and the antidot potential (Eq.~(\ref{potenz})) 
with $\beta=2$.  For $p/q=1/2$ the Hall conductances in the minigaps are shown and also for $p /q=,1/3,$ and $2/3$, if they  deviate from the corresponding value of Hofstadter's butterfly (given in brackets). 
21 Landau bands were numerically taken into account.}
\end{minipage}

\end{figure*}
\clearpage

Restricting ourself to the case of non-overlapping Landau bands and $p/q <1$, we find for different potential shapes and strengths, that the largest minigaps  usually occur at $p/q \approx 1/2, 1/3,$ and  $2/3$.  
The corresponding Hall conductances will thus  be the first to be found
experimentally. They will differ from the Hall conductances of  Hofstadter's
butterfly, as discussed in the last section and shown in Fig.~5. 

\section{conclusion}

We have studied the energy spectrum of electrons in a two-dimensional periodic
potential with perpendicular magnetic field {\it without} neglecting  the coupling of the Landau bands. 
We examined the Hall conductance, since its values, which 
are quantized in every energy gap, 
contain quantitative information about the
structure of the spectrum. 
The Landau band coupling changes this structure compared to Hofstadter's
butterfly, resulting in dramatical  modifications of the Hall conductance. 
We find the following three principal deviations from the Hall conductance in  Hofstadter's butterfly: i) opening of previously closed gaps, ii) rearrangement of
subbands, including their contributions to the Hall conductance, and iii) unexpected subband contributions to the Hall conductance.
Remarkably, even for weak Landau band coupling these changes can be found. 
This was explained by the occurrence of crossings of branches of the
spectrum, e.g.\ flatband positions, which are very sensitive to the Landau
band coupling. 
\acknowledgments

This work was supported by the Deutsche Forschungsgemeinschaft.


\begin{references}
\bibitem{zz}
In order to find the extrema of the  energy subbands we solve Eq.~(\ref{schr}) numerically by standard diagonalization techniques 
for a number of  combinations of the phases $\kappa$ and $\vartheta$ making use of
symmetries.
\end{references}
\end{document}